\documentclass[11pt]{article}

\usepackage{hyperref}
\usepackage[bf]{caption}
\usepackage{epsfig}
\usepackage{float}
\usepackage{graphicx}
\usepackage{subfig}
\usepackage{color}
\def\ga{\mathrel{\raise.3ex\hbox{$>$\kern-.75em\lower1ex\hbox{$\sim$}}}}
\def\la{\mathrel{\raise.3ex\hbox{$<$\kern-.75em\lower1ex\hbox{$\sim$}}}}

\def\I_M{{I_{\scriptscriptstyle M\times M}}}

\begin{document}

\begin{titlepage}

\thispagestyle{empty}

\vskip 1.2cm

\begin{centering}
\vspace{1cm}
{\Large {\bf Generalized Holographic Cosmology}}\\

\vspace{1.5cm}
{\bf Souvik Banerjee $^{\dagger}$}, {\bf Samrat Bhowmick $^{\dagger}$},\\
{\bf Anurag Sahay $^{\dagger}$}  \\
 \vspace{.1in}
 Institute of Physics \\
Bhubaneswar -751 005, India \\
\vspace{.5in}
 {\bf George Siopsis $^{\flat}$}
\vspace{.1in}

 Department of Physics and Astronomy, The
University of Tennessee,\\ Knoxville, TN 37996 - 1200, USA
 \\
\vspace{3mm}

\end{centering}
\vspace{1.5cm}

\vskip 1.2cm
\vskip 1.2cm
\centerline{\bf Abstract}
\noindent
We consider general black hole solutions in five-dimensional spacetime in the presence of a negative cosmological constant. We obtain a cosmological evolution via the gravity/gauge theory duality (holography) by defining appropriate boundary conditions on a four-dimensional boundary hypersurface. The standard counterterms are shown to renormalize the bare parameters of the system (the four-dimensional Newton's constant and cosmological constant). We discuss the thermodynamics of cosmological evolution and present various examples. The standard brane-world scenarios are shown to be special cases of our holographic construction.

\vspace{3.5cm}

\begin{flushleft}
$^{\dagger}~~$ e-mail address: souvik,samrat,anurag@iopb.res.in \\
$ ^{\flat}~~$ e-mail address: siopsis@tennessee.edu

\end{flushleft}
\end{titlepage}
\newpage
\setcounter{footnote}{0}
\noindent

\section{Introduction}
In recent years, understanding cosmology within the framework of string theory, has been an
active and interesting field of study. Starting with \cite{rs}, a substantial amount of research has been based on modeling the Universe by a 3-brane living in a higher-dimensional bulk space (brane world scenario). An incomplete list of references is \cite{binetruy,bibn1,bibn2,bibn3,Kraus:1999it,bibn4,bibn5,apostol,exchange,bibn7,bibn8,bibn9}. The Hubble equation of cosmological evolution is thus reproduced by the trajectory of the brane.

A different approach to understanding dynamics starting with a higher-dimensional bulk space is provided by the AdS/CFT correspondence \cite{bib1}. In this approach, a solution of the Einstein equations in a space with a negative cosmological constant is shown to be dual to a gauge theory of lower dimension living on the boundary (gravity/gauge theory duality). The stress-energy tensor of the gauge theory is constructed using holography \cite{Skenderis}. This affords an understanding of the gauge theory at strong coupling, which has been applied to various physical systems, such as the quark-gluon plasma formed in heavy ion collisions and high temperature superconductors. It is possible to understand the motion of a fluid, including time-dependent flows. An application to cosmology, which is our interest here, has also been discussed \cite{bibn6,Apostolopoulos:2008ru}.

Applying the gravity/gauge theory duality to a cosmological setting is not straightforward due to the fact that the metric on the boundary space in which the gauge theory lives must remain dynamical. This was long thought to be problematic due to the possibility of the fluctuations of the bulk metric corresponding to non-normalizable modes \cite{bib2,bib2a,bib2b,bib2c,bib2d,bib2e,bib2f}.
It was shown in \cite{Compere:2008us}, that such problems can be avoided by introducing appropriate counterterms on the boundary needed to cancel the infinities.

In applications of the gravity/gauge theory duality (holography) to cosmology and other settings, one generally places the boundary at a finite distance $r$ and then takes the limit as the cutoff $r\to\infty$. The removal of the cutoff introduces infinities, which are canceled by the addition of a local action on the boundary with $r$-dependent coefficients (counterterms). Unlike in quantum field theory, where counterterms are interpreted as renormalization of the (bare) parameters of the system, it is not clear if counterterms have a similiar physical meaning in a holographic setting.

Here we generalize the holographic approach to cosmology by placing the boundary hypersurface at a finite distance $r$ and derive expressions for the various physical quantities (e.g., the stress-energy tensor) which are valid for arbitrary $r$. This leads to a generalized Hubble equation of cosmological evolution. We still need to introduce the standard counterterms to avoid infinities at large $r$. We show that these counterterms have the usual field theoretic interpretation of renormalizing the (bare) parameters of the system, namely Newton's constant and the cosmological constant.
Moreover, we recover the brane-world scenario by fine-tuning Newton's constant. Thus we show that brane-world scenarios are a special case of our generalized holographic approach.
 
Our paper is organized as follows. In section \ref{sec:2} we discuss the bulk space concentrating on a time-independent solution (general black hole) of the field equations, and define the boundary hypersurface. In section \ref{sec:3} we introduce the boundary conditions and the counterterms needed to cancel infinities. We calculate the stress-energy tensor and derive the Hubble equation of cosmological evolution. In section \ref{sec:4} we discuss the example of a bulk Reissner-Nordstr\"om black hole including thermodynamics. In section \ref{sec:5} we discuss various examples of cosmological evolution. In particular, we show that the brane-world  scenarios are a special case of our holographic approach. Finally in section \ref{sec:6} we conclude.

\section{The Bulk}\label{sec:2}
We start with a non-extremal black hole in a $4+1$ dimensional bulk space in the presence of a negative cosmological constant
\begin{equation}\label{eqL5} \Lambda_5 = - \frac{6}{L^2}~. \end{equation}
We consider the metric \emph{ansatz}
\begin{equation}
\label{metric5}
 ds_5^2 = - A(r) dt^2 + B(r) dr^2 + r^2 d\Omega_k^2 \;,
\end{equation}
$r$ being the radial direction, and $k =+1, 0 , -1$ depending on the geometry of the constant $(t,r)$ hypersurfaces (spherical, flat, or hyperbolic, respectively). More general metrics are also possible, but will clutter the notation unnecessarily. In section \ref{sec:4}, we shall concentrate on the special case of a Reissner-Nordstr\"om black hole for explicit calculations.

Asymptotically, we have AdS space of radius $L$, therefore as $r\to\infty$,
\begin{equation} A(r)\approx \frac{1}{B(r)} \approx \frac{r^2}{L^2} ~.\end{equation}
We introduce a radial cutoff, $r=a$ and 
parametrise $a$ and $t$ as $a = a(\tau)$ and $t= t(\tau)$ so that $a = \dot{a} dt$.
Then the metric on the cut-off surface (boundary) takes the form
\begin{equation}
 ds_4^2 = \left[-A(a) (\frac{dt}{d\tau})^2 + B(a) \dot{a}^2\right] d\tau^2 + a^2(\tau) d\Omega_k^2 \;.
\end{equation}
In order that the metric on the boundary take the FRW form,
\begin{equation}
\label{FRW}
 ds_4^2 = -d\tau^2 + a^2(\tau) d\Omega_k^2~,
\end{equation}
the metric components should satisfy the relation
\begin{equation}\label{eqB}
 \left( \frac{dt}{d\tau} \right)^2  = \frac{\mathcal{B} }{A} \ \ , \ \ \ \ \mathcal{B} = B(a) \dot{a}^2 +1 \;.
\end{equation}
This in turn fixes our choice of the time parameter $\tau$. Notice also that if $T_H$ is the Hawking temperature, then the temperature on the boundary is redshifted,
\begin{equation}\label{eqT}
T = \frac{T_H}{\sqrt{A\mathcal{B}}} ~. \end{equation}
This kind of parametrization has been used before, e.g., in
\cite{Barcelo:2000re,Savonije:2001nd,Cardoso:2008gm}. Note that, while treating 
$\tau$ as a time parameter, we are effectively considering the radial motion of the cut-off surface in 
the $4+1$ dimensional bulk. By adopting appropriate boundary conditions, the cut-off surface can be thought of as the location of a brane, 
mimicking a moving brane scenario.

\section{Boundary Conditions}\label{sec:3}

The heart of the construction we are going to elaborate on is based on the observation that 
the afore-mentioned dynamics of the boundary hypersurface will be captured through the boundary conditions
we impose on the system. This approach was first adopted in \cite{Apostolopoulos:2008ru}. 

Let us consider a general five dimensional bulk action,
\begin{equation} S_5 = \int_{\mathcal{M}} d^5 x \sqrt{-g} \mathcal{L}_5 \; , \end{equation}
where we keep the Lagrangian density $\mathcal{L}_5$ 
unspecified. In the simplest case, this consists of a five-dimensional Einstein-Hilbert action with a 
negative cosmological constant (\ref{eqL5}) plus the requisite Gibbons-Hawking surface term for a well-defined variational principle. If one varies this action with respect to the metric, one obtains a
boundary term of the form
\begin{equation}
\label{var1}
\delta S_5 = \frac{1}{2} \int_{\partial\mathcal{M}} d^4x \sqrt{-\gamma} T_{\mu\nu}^{\mathrm{(CFT)}} \delta \gamma^{\mu\nu},
\end{equation}
where $\gamma^{\mu\nu}$ is the induced boundary metric and $\gamma$ is its determinant.
$T_{\mu\nu}^{\mathrm{(CFT)}}$ denotes the (bare) stress-energy tensor of the dual conformal field theory that lives on 
the four-dimensional boundary hypersurface $r=a$. Generally in the context of the AdS/CFT correspondence, Dirichlet boundary 
conditions are employed, which fix the boundary metric and consequently eq.~(\ref{var1}) vanishes. While this leads to a well-defined variational principle, it does not allow for a dynamical boundary metric.
Since we are primarily interested in obtaining a cosmological evolution and hence a dynamical metric on the
boundary, we seek different boundary conditions that can be imposed without fixing the metric on the 
boundary. It was noted in \cite{Apostolopoulos:2008ru} that one could adopt appropriate \emph{mixed} boundary conditions, which were shown to lead to valid dynamics in \cite{Compere:2008us}. Their definition involves the addition of an appropriate local action, $S_{\mathrm{local}}$, at the boundary.
For cosmological evolution, this local action will be chosen as the four-dimensional Einstein-Hilbert action on the boundary with
an arbitrary (positive, negative, or vanishing) four-dimensional cosmological constant $\Lambda_4$,
\begin{equation}
\label{Slocal}
 S_{\mathrm{local}} = -\frac{1}{16 \pi G_4}\int_{\partial\mathcal{M}} d^4 x \sqrt{-\gamma}(\mathcal{R[\gamma]}-2 \Lambda_4),
\end{equation}
where $\mathcal{R[\gamma]}$ is the Ricci scalar evaluated with the boundary metric which, in our case, is the 
FRW metric (\ref{FRW}). Notice that the cosmological constant may be due, wholly or partly, to a brane of finite tension at the boundary.

Additionally, to cancel divergences in the limit $a\to\infty$, it is necessary to introduce counterterms \cite{Skenderis}. These are of the same form as the local action and renormalize the four-dimensional physical parameters $G_4$ and $\Lambda_4$. We have
\begin{equation}
\label{Sct}
 S_{\mathrm{c.t.}} = -\frac{1}{2}\int_{\partial\mathcal{M}} d^4 x \sqrt{-\gamma}\left( \kappa_1 \mathcal{R[\gamma]}+ \kappa_2 \right) ~,
\end{equation}
which diverges as $a\to\infty$.
The parameters $\kappa_1$ and $\kappa_2$ will be chosen so that physical quantities such as the energy density and pressure remain finite in this limit.

Putting these pieces together, we \emph{define} our boundary condition as
\begin{equation}
\label{bdycondT}
T_{\mu\nu}^{\mathrm{(CFT)}}+ T_{\mu\nu}^{\mathrm{(local)}}+T_{\mu\nu}^{\mathrm{(c.t.)}}=0,
\end{equation}
where $T_{\mu\nu}^{\mathrm{(CFT)}}$ is due to the variation $\delta S_5$ (eq.~(\ref{var1})), and the other two terms, $T_{\mu\nu}^{\mathrm{(local)}}$ and $T_{\mu\nu}^{\mathrm{(c.t.)}}$ come from the variations
\begin{eqnarray}
\delta S_{\mathrm{local}} &=& \frac{1}{2} \int_{\partial\mathcal{M}} d^4x \sqrt{-\gamma} T_{\mu\nu}^{\mathrm{(local)}} \delta \gamma^{\mu\nu}\; , \nonumber\\
\delta S_{\mathrm{c.t.}} &=& \frac{1}{2} \int_{\partial\mathcal{M}} d^4x \sqrt{-\gamma} T_{\mu\nu}^{\mathrm{(c.t.)}} \delta \gamma^{\mu\nu} \; , \end{eqnarray}
respectively, with respect to the boundary metric, $\gamma_{\mu\nu}$.
Similarly to Dirichlet boundary conditions, the choice (\ref{bdycondT}) leads to a well-defined variational principle with
\begin{equation}
\label{bdycond}
\delta S_5+ \delta S_{\mathrm{local}}+\delta S_{\mathrm{c.t.}}=0 \; .
\end{equation}
To see the explicit physical content of our \emph{mixed} boundary conditions (\ref{bdycondT}), we shall derive explicit expressions for each of the three contributing terms. The bare stress-energy tensor on the boundary is given by 
\begin{equation}
 \label{Stress-energy-bare}
T_{\mu\nu}^{\mathrm{(CFT)}} = \frac{1}{8 \pi G_5}(\mathcal{K}_{\mu\nu}-\mathcal{K} \gamma_{\mu\nu})\;,
\end{equation}
where $\mathcal{K}_{\mu\nu}$ is the extrinsic curvature, and $\mathcal{K}$ is its trace.
The components of this tensor can be evaluated by computing the velocity $v^\mu$ and unit normal $n^\nu$ vectors on
the boundary hypersurface, $r=a(\tau)$. For the metric (\ref{metric5}), these vectors are given in 
component form as
\begin{equation}
 v^\mu=\left( \sqrt{\frac{\mathcal{B}}{A}},\dot{a},0,0,0 \right)\; ;
~~~~~~~~~~
v_\mu=\left( -\sqrt{A\mathcal{B}},B\dot{a},0,0,0 \right)\; .
\end{equation}
and
\begin{equation}
 n^\mu = \left(- \sqrt{\frac{B}{A}}\dot{a},
-\sqrt{\frac{\mathcal{B}}{B}},0,0,0\right) \; ;  ~~~~~~~~~~~
 n_\mu = \left( \sqrt{AB}\,\dot{a},
-\sqrt{B\mathcal{B}},0,0,0\right) \;,
\end{equation}
respectively.
The direction of the unit normal vector is taken to be pointing inward, toward the bulk.
The extrinsic curvature can be written in terms of the unit normal and velocity vectors as
\begin{equation}
\mathcal{K}_{ij} = \frac{1}{2} n^{k}\partial_{k} \gamma_{ij}  ~~~~~~~~~~~~~~
\mathcal{K}_{\tau\tau} = -\frac{\partial_{\tau} v_t}{n_t}.
\end{equation}
Explicitly, they are
\begin{equation}
\label{excurv}
 \mathcal{K}_{ij} =
a \sqrt{\frac{\mathcal{B}}{B}} \gamma_{ij}\;, ~~~~~~~~~~~~~~
\mathcal{K}_{\tau\tau} = -\frac{3}{2aAB}
\left( 2 A B \ddot{a}+ (AB)' \dot{a}+A' \right) \;,
\end{equation}
where $i,j$ are indices for the spatial coordinates on the boundary (spanned by $\Omega_k$).

We deduce the explicit expressions for the components of the bare stress-energy tensor
 (\ref{Stress-energy-bare}),
\begin{eqnarray}
\label{se1}
 T_{\tau\tau}^{\mathrm{(CFT)}} &=& -\frac{3}{8 \pi G_5 a} \sqrt{\frac{ \mathcal{B}}{B}} \; , \\
 T_{i}^{i\, \mathrm{(CFT)}} &=& \frac{1}{16 \pi G_5}\frac{a A' \mathcal{B}+A
   [ a B' \dot{a}+2 B \left(a \ddot{a}+2 \dot{a}^2\right)
+4]}{aA \sqrt{B\mathcal{B}}} \; ,
\end{eqnarray}
where no summing over the index $i$ is implied. Notice that the energy density $T_{\tau\tau}^{\mathrm{(CFT)}}$ obtained above is negative, however we should emphasize that this is only a \emph{bare} quantity and therefore not physical. It will be corrected by the addition of counter terms resulting into a \emph{positive} regularized (physical) quantity.

For the remaining two contributions in (\ref{bdycondT}), we obtain the standard expressions one encounters in Einstein's four-dimensional equations,
\begin{equation}
T_{\mu\nu}^{\mathrm{(local)}} = -\frac{1}{8\pi G_4} \left( \mathcal{R}_{\mu\nu} - \frac{1}{2}\gamma_{\mu\nu} \mathcal{R} - \Lambda_4 \gamma_{\mu\nu} \right) \; , ~~~~~~~~
T_{\mu\nu}^{\mathrm{(c.t.)}} = -\kappa_1 \left( \mathcal{R}_{\mu\nu} - \frac{1}{2}\gamma_{\mu\nu} \mathcal{R} \right) - \kappa_2 \gamma_{\mu\nu} \; ,
\end{equation}
where $\mathcal{R}_{\mu\nu}$ ($\mathcal{R}$) is the four-dimensional Ricci tensor (scalar) constructed from the four-dimensional boundary metric $\gamma_{\mu\nu}$. The counter terms diverge in the limit $a\to\infty$, and the parameters $\kappa_1$ and $\kappa_2$ will be chosen so that they cancel the divergences in the bare stress-energy tensor $T_{\mu\nu}^{\mathrm{(CFT)}}$. Notice that the counter terms  are of the same form as the terms coming from the local action. Therefore, they admit the standard interpretation of inducing the renormalization of the physical four-dimensional constants $G_4$ (Newton's constant) and $\Lambda_4$ (cosmological constant).

The regularized (physical) stress-energy tensor is
\begin{equation}
\label{ge}
T^{\mathrm{(reg)}}_{\mu\nu} =T^{\mathrm{(CFT)}}_{\mu\nu} + T^{\mathrm{(c.t.)}}_{\mu\nu} \; .
\end{equation}
We deduce the energy density and pressure, respectively,
\begin{eqnarray}
\label{se1ph}
 \epsilon = T_{\tau\tau}^{\mathrm{(reg)}} &=& \kappa_2 + \kappa_1 \left(H^2+\frac{k}{a^2}\right)-\frac{3}{8 \pi G_5 a} \sqrt{\frac{ \mathcal{B}}{B}} \; , \nonumber\\
 p = T_{i}^{i\, \mathrm{(reg)}} &=& -\kappa_2 -\kappa_1
\left\{\left(H^2+\frac{k}{a^2}+\frac{2\ddot{a}}{a}\right)\right\} +\frac{1}{16 \pi G_5}\frac{a A' \mathcal{B}+A
   [ a B' \dot{a}+2 B \left(a \ddot{a}+2 \dot{a}^2\right)
+4]}{aA \sqrt{B\mathcal{B}}} \; .\nonumber\\
\end{eqnarray}
where $H =\dot{a}/a$ is the Hubble parameter.
The choice
\begin{equation}\label{eqchoice} \kappa_1 = \frac{3L}{ 16\pi G_5 } \ \ , \ \ \ \ \kappa_2 = \frac{3}{8\pi G_5L} \end{equation}
ensures finiteness in the limit $a\to\infty$.
Unlike the bare energy density (\ref{se1}), the regularized energy density $\epsilon$ is positive.

The boundary conditions (\ref{bdycondT}) now read
\begin{equation}
 \label{eveq}
\mathcal{R}_{\mu\nu} - \frac{1}{2} \gamma_{\mu\nu}\mathcal{R} -\Lambda_4 \gamma_{\mu\nu}
= 8\pi G_4 T_{\mu\nu}^{\mathrm{(reg)}}~,
\end{equation}
which are the four-dimensional Einstein equations in the presence of a cosmological constant.

The cosmological evolution equation is the $\tau\tau$ component of the Einstein equations (\ref{eveq}),
\begin{equation}
\label{evequnren}
H^2+\frac{k}{a^2}-\frac{\Lambda_4}{3} = \frac{8\pi G_4}{3} \epsilon ~,
\end{equation}
where $\epsilon$ is the energy density given in (\ref{se1ph}) under the condition (\ref{eqchoice}). This is deceptively similar to the standard equation of cosmological evolution. However, it differs in an essential way, because $\epsilon$ contains contributions that involve the Hubble parameter $H = \dot{a}/a$, leading to novel cosmological scenarios.

\section{AdS Reissner-Nordstr\"{o}m black hole}\label{sec:4}
In this section we take up the example of an asymptotically AdS charged black hole,
namely AdS Reissner-Nordstr\"{o}m black hole for which the functions $A$ and $B$ of (\ref{metric5}) are 
\begin{equation}
A(r) = \frac{1}{B(r)} = \frac{r^2}{L^2}+k-\frac{M}{r^2}+\frac{Q^2}{r^4} ~. \end{equation}
The parameters $M$ and $Q$ are related to
the mass and charge of the black hole, respectively. $k$ can be $+1$, $0$, or $-1$ depending on whether the
black hole horizon is spherical, flat, or hyperbolic, respectively.

The Hawking temperature is
\begin{equation} T_H = \frac{2\frac{r_+^2}{L^2} + k}{2\pi r_+} ~, \end{equation}
where $r_+$ is the radius of the horizon satisfying
\begin{equation}\label{eqrh} A(r_+) = \frac{r_+^2}{L^2}+k-\frac{M}{r_+^2}+\frac{Q^2}{r_+^4} = 0 ~. \end{equation}
The entropy is
\begin{equation} S = \frac{r_+^3}{4G_5} V_3 ~, \end{equation}
where $V_3$ is the three-dimensional volume spanned by $\Omega_k$. Notice that the entropy is independent of $a$, and therefore
constant in time, leading to an adiabatic evolution.

According to (\ref{eqT}), the redshifted temperature on the boundary is
\begin{equation} T = \frac{T_H}{\sqrt{\dot{a}^2 + A(a)}} ~. \end{equation}
For large $a$, it is expanded as
\begin{equation} T = \frac{T_H L}{a} - \frac{T_HL^3}{2a} \left( H^2 + \frac{k}{a^2} \right)+ \dots~. \end{equation}
Similarly, we expand the regularized energy density and pressure (\ref{se1ph}), respectively,
\begin{eqnarray}
\label{exptaureg}
 \epsilon &=&
\frac{3L^3}{64\pi G_5} \left\{\left(H^2+\frac{k}{a^2}\right)^2
+\frac{4M}{L^2 a^4}\right\} \nonumber\\
&& - \frac{3L^5}{128\pi G_5} \left\{\left(H^2+\frac{k}{a^2}\right)^3+\frac{4kM}{L^2a^6}+
\frac{8Q^2}{L^4a^6}+\frac{4M}{L^2}\frac{H^2}{a^4}\right\} + \dots ~, \\
\label{expireg}
p &=&  \frac{L^3}{64\pi G_5} 
\left\{\left(H^2+\frac{k}{a^2}\right)^2+\frac{4M}{L^2 a^4}
-4\left(H^2+\frac{k}{a^2}\right)\frac{\ddot{a}}{a}\right\} \nonumber\\
&& - \frac{3L^5}{128\pi G_5}\left\{\left(H^2+\frac{k}{a^2}\right)^3+\frac{4kM}{L^2a^6} +
\frac{8Q^2}{L^4a^6}+\frac{4M}{L^2}\frac{H^2}{a^4}
 - 2\left(H^2+\frac{k}{a^2}\right)^2 \frac{\ddot{a}}{a} 
- \frac{8}{3L} \frac{M \ddot{a}}{a^5} \right\}\nonumber\\
&& +\dots~.
\end{eqnarray}
We deduce the conformal anomaly which is given by the trace of the stress-energy tensor,
\begin{eqnarray}
\mathrm{Tr} T &=& \epsilon -3p \nonumber\\
&=&
- \frac{3L^3}{16\pi G_5} \left(H^2+\frac{1}{a^2}\right)\frac{\ddot{a}}{a} \nonumber \\
&& - \frac{3L^5}{64\pi G_5}  \left\{\left(H^2+\frac{k}{a^2}\right)^3+\frac{4kM}{L^2a^6}+
\frac{8Q^2}{L^4a^6}+\frac{4M}{L^2}\frac{H^2}{a^4}
-3 \left(H^2+\frac{k}{a^2}\right)^2 \frac{\ddot{a}}{a}
-\frac{4}{L}\frac{M\ddot{a}}{a^5}\right\}\nonumber\\
&& +\dots \;.
\end{eqnarray}
The first term is the standard conformal anomaly one obtains in the large $a$ limit \cite{Apostolopoulos:2008ru}.

As an example, consider the case of a flat static boundary of a Schwarzschild black hole. Then $k=0$, $Q=0$, and $H=0$.
The radius of the horizon is $r_+ = (ML^2)^{1/4}$. The expressions for the energy density, pressure and temperature simplify to, respectively,
\begin{equation} \epsilon = T_{\tau\tau} = \frac{3}{8\pi G_5 L} \left( 1 - \sqrt{1 - \frac{r_+^4}{a^4}} \right) \ \ , \ \ \ \
p = T_i^i = \frac{1}{8\pi G_5 L}\left( \frac{3-\frac{r_+^4}{a^4}}{ \sqrt{1- \frac{r_+^4}{a^4}}} -3  \right) \ \ , \ \ \ \ 
T = \frac{r_+}{\pi L a \sqrt{1- \frac{r_+^4}{a^4}}} \end{equation}
In the large $a$ limit, we deduce the expansions
\begin{equation} \epsilon = \frac{3}{8\pi G_5 L} \left( \frac{(\pi LT)^4}{2} - \frac{7(\pi LT)^8}{8} + \dots \right) \ \ , \ \ \ \
p = \frac{1}{8\pi G_5 L} \left( \frac{(\pi LT)^4}{2} - \frac{3(\pi LT)^8}{8} + \dots \right) ~. \end{equation}
Thus, at leading order, we have $\epsilon = 3p \propto T^4$, as expected for a conformal fluid. Including next-order corrections, we no longer have a traceless stress-energy tensor.

Returning to the general case, we obtain the law of thermodynamics
\begin{equation} dE = TdS - pdV + \Phi dQ ~, \end{equation}
where $E=\epsilon V$, $V= a^3 V_3$ is the volume, and $\Phi$ is the potential
\begin{equation} \Phi = \frac{Q}{G_5 a}~. \end{equation}
This is easily verified, e.g., by differentiating with respect to $\tau$, $r_+$, and $Q$ (after using (\ref{eqrh}) to express $M$ in terms of the other two parameters, $r_+$ and $Q$).


\section{Cosmological Evolution}\label{sec:5}

Next, we discuss various explicit examples of cosmological evolution based on an AdS Reissner-Nordstr\"om black hole.
For simplicity, in what follows we shall be working with units in which $L=1$.

The Hubble equation (\ref{evequnren}) can be massaged into the form
\begin{equation}\label{V1a}
\beta \left( H^2 + \frac{k}{a^2} \right) = \frac{1}{L'} - \sqrt{ H^2 + \frac{A(a)}{a^2} }  ~, \end{equation}
where we introduced the convenient combinations of parameters
\begin{equation}\label{V1ap} \beta = \frac{G_5}{G_4} - \frac{1}{2} \ , \ \ \ \frac{1}{L'} = 1 + \frac{(1+ 2\beta )\Lambda_4}{6} \; .
\end{equation}
The Hubble equation can be expanded for large $a$ as
\begin{eqnarray}
\label{hubeq}
H^2  + \frac{k}{a^2} - \frac{\Lambda_4}{3} &=& \frac{G_4L^3}{16G_5} \left\{ \left(H^2 + \frac{1}{a^2}\right)^2+
 \frac{4M}{L^2a^4}\right\} \nonumber\\
& & - \frac{G_4L^4}{16G_5} \left\{ \left(H^2 + \frac{1}{a^2}\right)^3  + \frac{4M}{L^2a^6}
+\frac{8Q^2}{L^4a^6} + \frac{4M}{L^2 a^4} H^2 \right\} + \dots \;.
\end{eqnarray}
At leading order, it coincides with the result obtained in \cite{Apostolopoulos:2008ru}.

After squaring (\ref{V1a}), we obtain a quadratic equation for $H^2$. However, only one of the two roots is a solution of (\ref{V1a}). Let us concentrate on the range of parameters with $\beta > 0$, $L' >0$. We obtain
\begin{equation}\label{V1asol}
H^2 = \frac{\left(\frac{1}{L'} - \frac{k}{a^2} \beta \right)^2 - \frac{A(a)}{a^2} }{\frac{1}{2} + \frac{\beta}{L'} - \frac{k\beta^2}{a^2} + \sqrt{ \frac{1}{4} + \frac{\beta}{L'} +  (A(a)-k) \frac{\beta^2}{a^2}}} ~.
\end{equation}
This can be solved for $a=a(\tau)$ to obtain the orbit of the boundary hypersurface. Once a solution of (\ref{V1asol}) is obtained, we still need to verify that it satisfies (\ref{V1a}), because the solutions of (\ref{V1a}) in general form a subset of the solutions of (\ref{V1asol}).

The fixed points of the orbits are found by setting $H=0$ in (\ref{V1a}). They are solutions of
\begin{equation}\label{V1at}
V(a)\equiv \frac{1}{L'} -\frac{k}{a^2}\beta - \frac{1}{a} \sqrt{ A(a)} =0 \; .
\end{equation}
These fixed points are also fixed points of (\ref{V1asol}), but the converse is not always true.

With the choice of parameters such that $\beta = 0$ \cite{Gubser:1999vj}, eq.\ (\ref{V1asol}) simplifies to
\begin{equation}\label{V1b}
H^2 = \left( 1 + \frac{\Lambda_4}{6} \right)^2 - \frac{A(a)}{a^2} ~,
\end{equation}
which coincides with the results from a brane world scenario.
Thus we recover the evolution of a 3-brane in a five-dimensional bulk space if we fine tune the parameters of our system so that $\beta =0$.

The fixed points are solutions of
\begin{equation}\label{V1bs}
V(a) \equiv 1+ \frac{\Lambda_4}{6} - \frac{1}{a} \sqrt{A(a)} =0 ~.
\end{equation}
Notice that no fixed points exist between the outer and inner horizons (with $A(a)<0$), because of the square root in the potential $V(a)$.
Notice also that $V(a) \approx \frac{\Lambda_4}{6}$ as $a\to\infty$, so the sign of the potential is determined by the sign of $\Lambda_4$, and at the horizon, $V(r_+) = \frac{1}{L'} > 0$. Up to two fixed points can be outside the horizon. However, our classical results likely receive significant quantum corrections as we approach the horizon. Therefore, our results are reliable for orbits away from the horizon, which typically end at infinite distance from the horizon.

\begin{figure}[H]
 \centering
 \subfloat[]{\label{fig:1a}
 \includegraphics[width=0.4\textwidth]{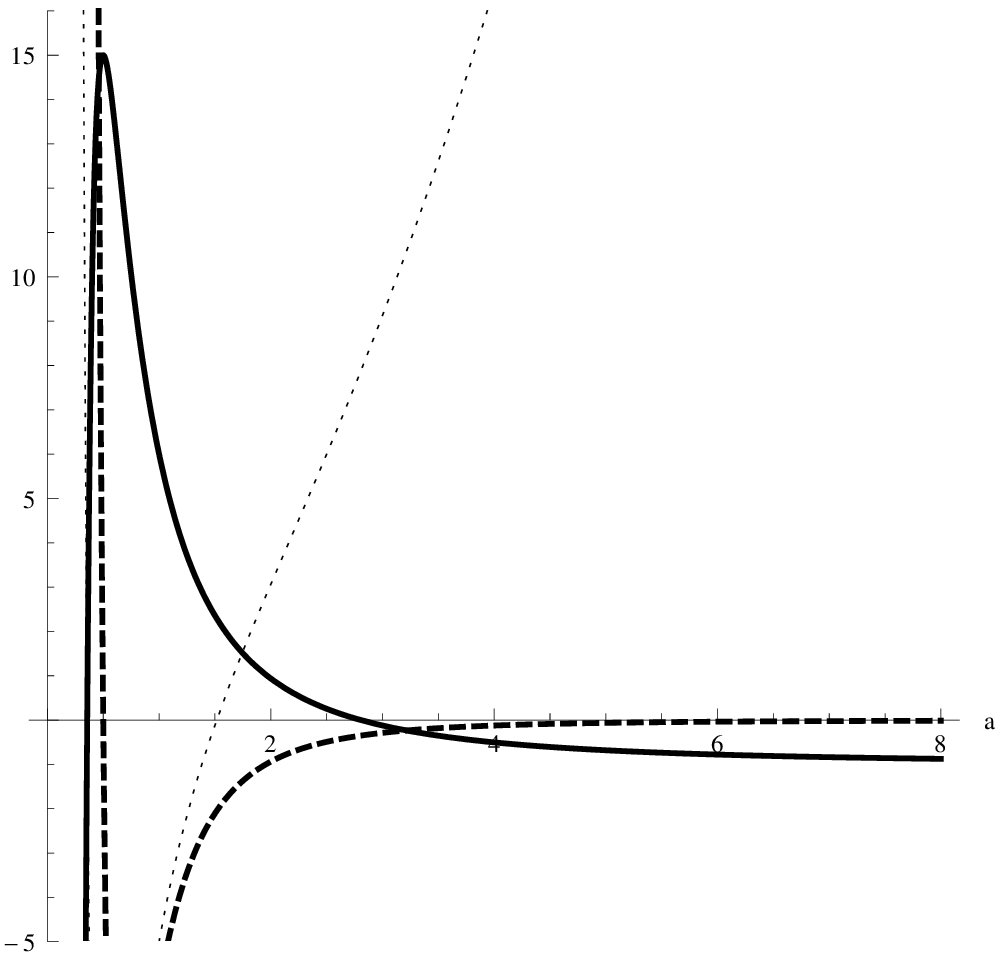}}
 ~~~~ ~~~~ 
 \subfloat[]{\label{fig:1b}
 \includegraphics[width=0.4\textwidth]{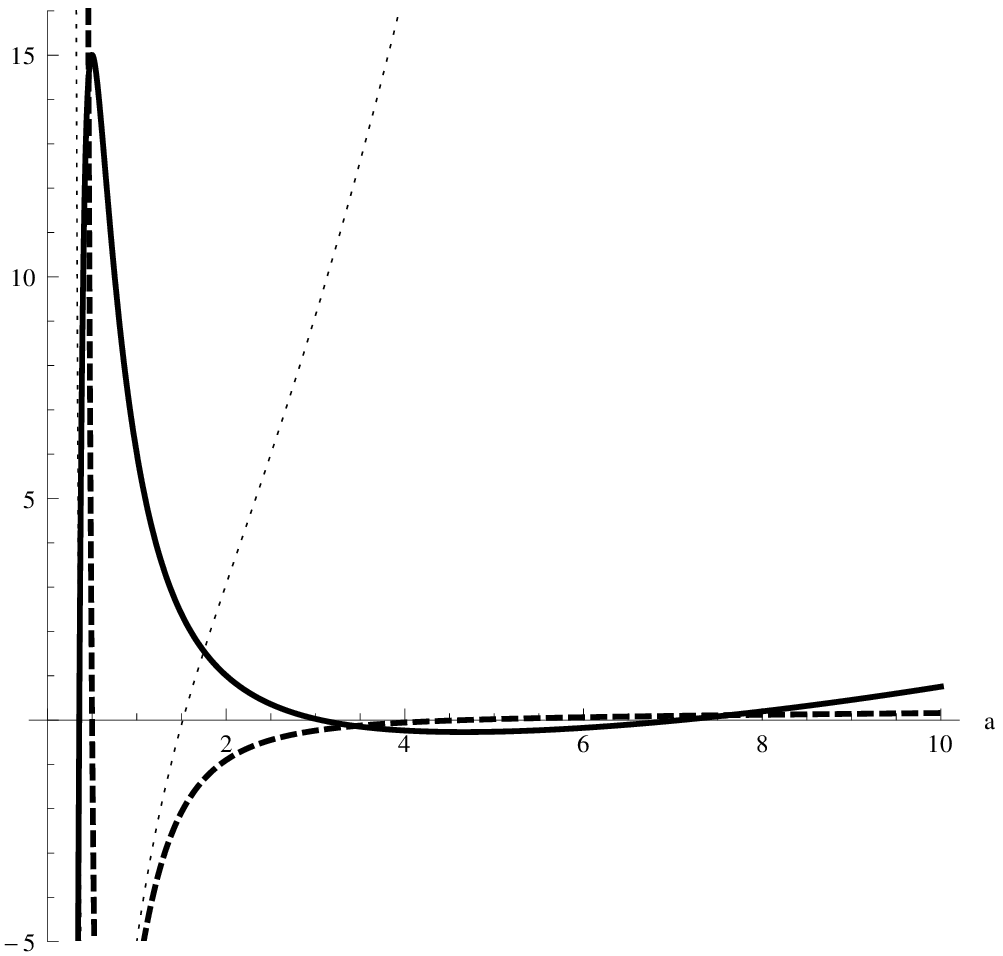}}

\subfloat[]{\label{fig:1c}
 \includegraphics[width=0.4\textwidth]{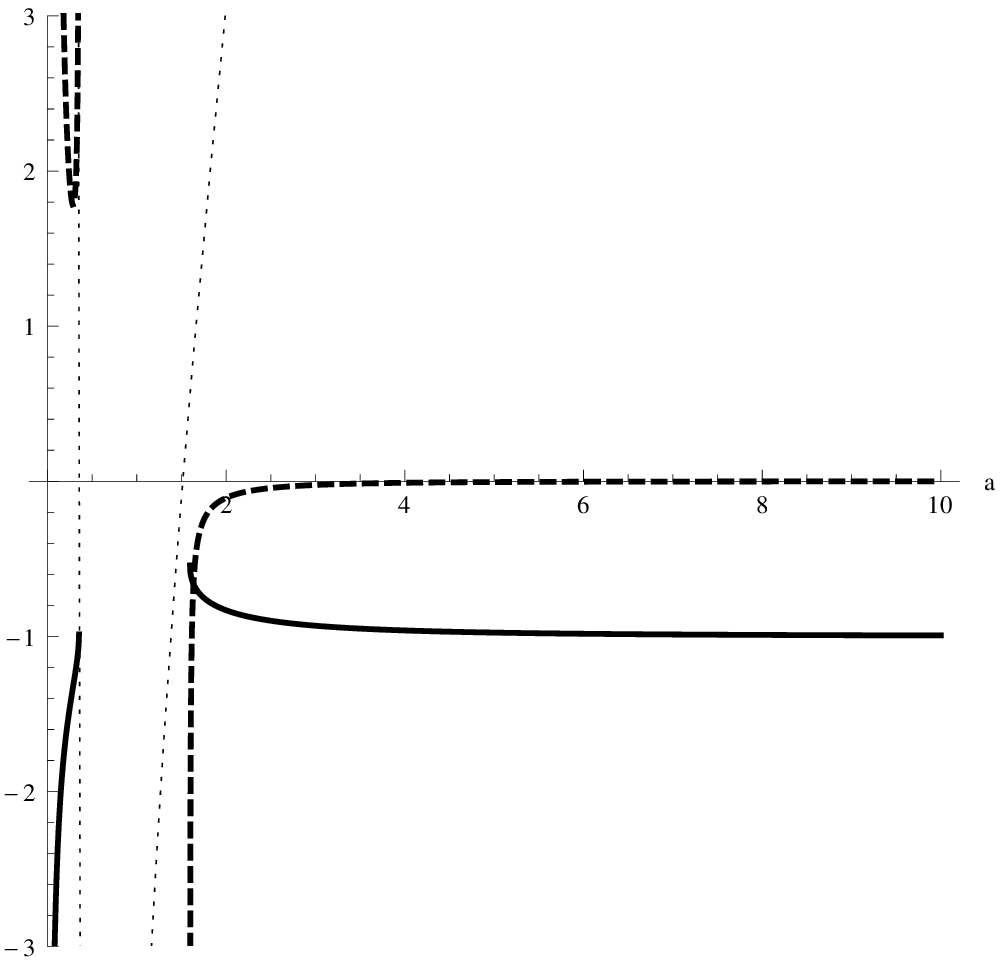}}
 ~~~~ ~~~~ 
 \subfloat[]{\label{fig:1d}
 \includegraphics[width=0.4\textwidth]{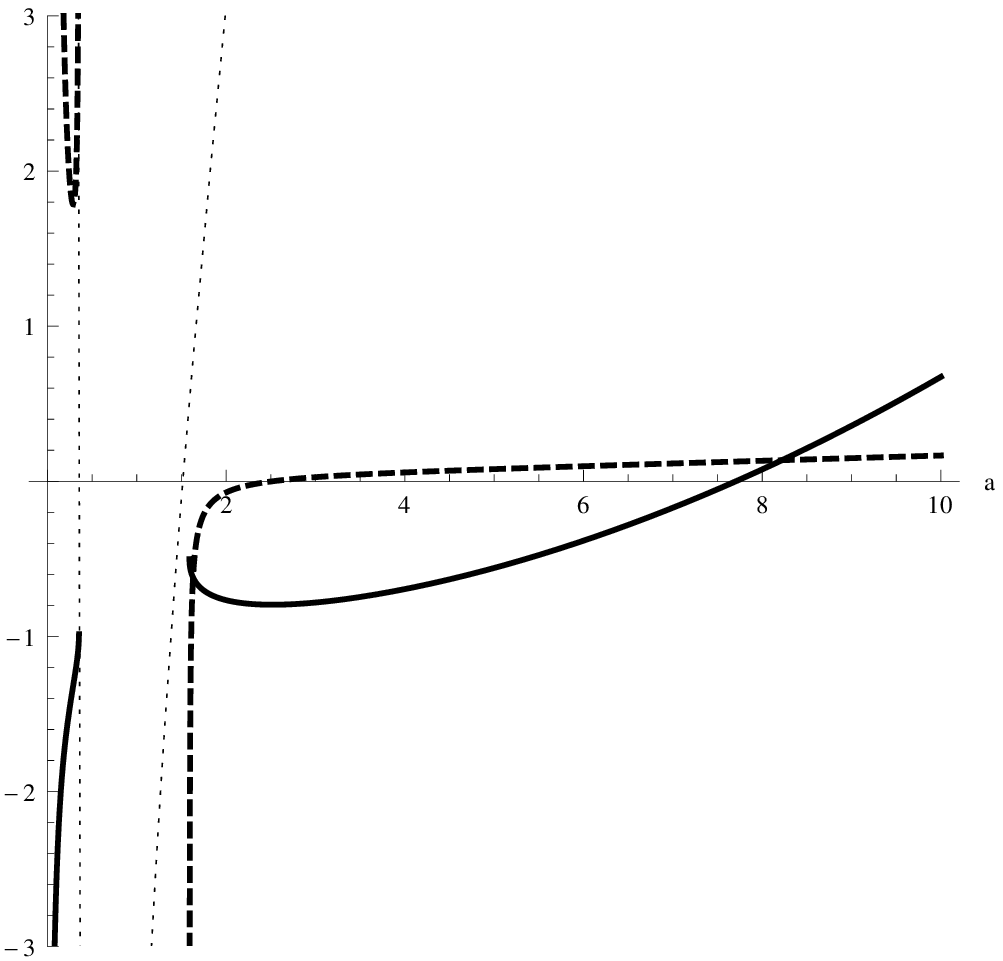}}

 \caption{Cosmological evolution scenarios for various values of parameters. Solid and dashed lines are plots of $\dot{a}^2$ and $\ddot{a}$, respectively.
Dotted lines denote the black hole potential with its zeros indicating the positions of
the inner and outer horizons.
}
\end{figure}

\begin{figure}[H]
 \centering

 \subfloat[]{\label{aZoom}
 \includegraphics[width=0.4\textwidth]{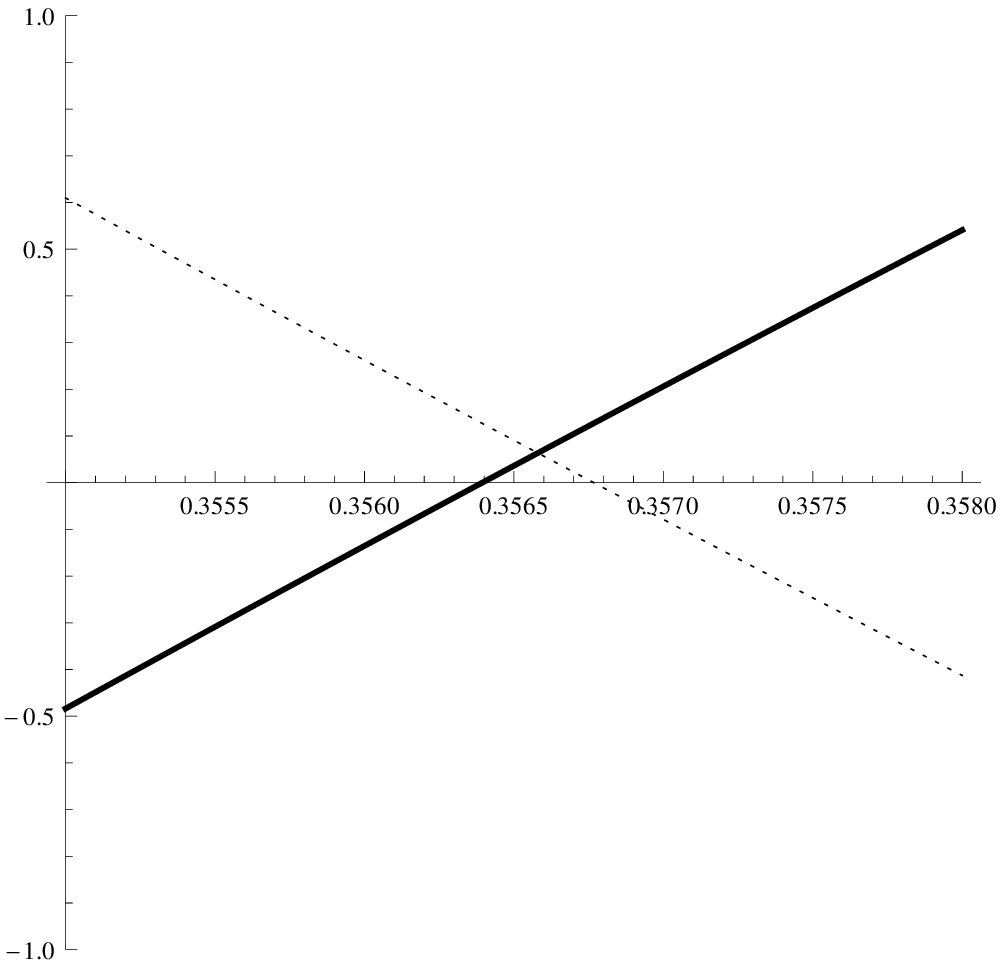}}
 ~~~~ ~~~~ 
 \subfloat[]{\label{bZoom}
 \includegraphics[width=0.4\textwidth]{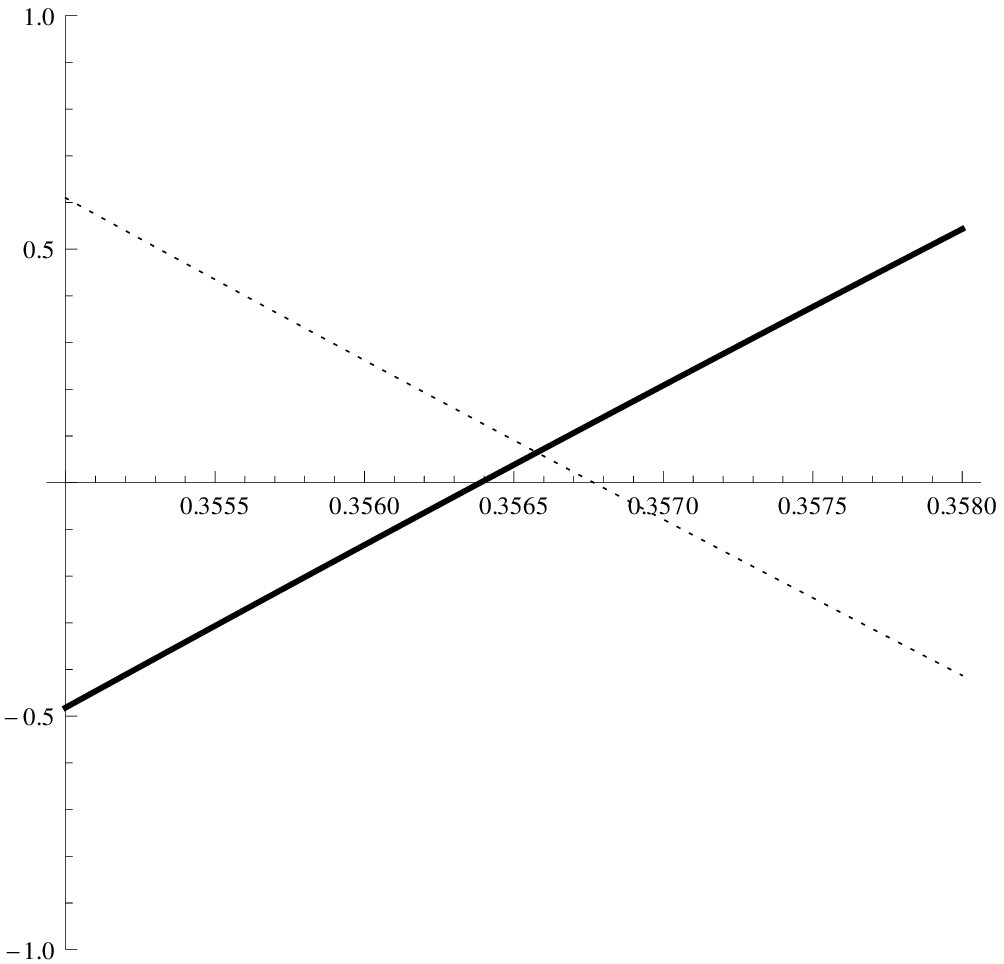}}

 \caption{Solid lines are plots of $\dot{a}^2$ whereas dotted lines 
are plots of the black hole potential for $\beta =0$ and (a) $\Lambda_4=0$, (b) $\Lambda_4 =0.05$.  The inner fixed points and the position of the inner horizon are shown.
}
\end{figure}

\begin{figure}[H]
 \centering
 \includegraphics[width=0.4\textwidth]{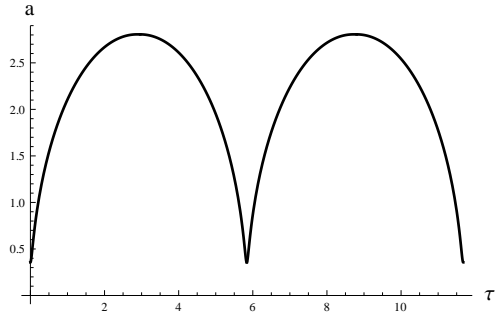}
 
\caption{\label{fig:00}Plot of $a$ vs $\tau$ for $\beta=0$, $\Lambda_4=0$.}
 
 \end{figure}

For $\Lambda_4 =0$, we recover from (\ref{V1b}) the brane world scenario of \cite{Mukherji:2002ft}.
This scenario is depicted in figure \ref{fig:1a} for $k=+1 \ , \ \ M = 8 \ , \ \ Q = 1$. 
We notice here that we have only one solution that is bouncing. Of the two 
turning points, one is inside the inner horizon and the other outside the outer horizon. 
There is no fixed point between the inner and outer horizons, as noted earlier, because of the presence of the 
square root in the potential $V(a)$ (\ref{V1at}). This can be explicitly seen
from figure \ref{aZoom} where we see clearly the position of the inner fixed point as the point 
where the solid line cuts the $a$-axis. After crossing the turning point outside the outer
horizon, the square of the Hubble parameter becomes negative and hence unphysical. 
The orbit of the bouncing solution is shown in figure \ref{fig:00}. 
Although we reproduce the bouncing cosmology of Mukherji, \emph{et al.}, through this, as argued in
\cite{Hovdebo:2003ug} this kind of solution suffers from an instability.
Indeed, the inner horizon is the Cauchy horizon for this charged AdS black hole and is 
unstable under linear fluctuations about the equilibrium black hole space-time. So when the orbit crosses the inner horizon of the black hole, it is not sufficient to consider only the 
unperturbed background. The backreaction on the background metric due to the fluctuating modes has 
to be taken into account. This backreaction is significant and may produce a curvature singularity. It should be noted that this pathology occurs only for $\beta =0$. For $\beta\ne 0$, no outward crossing of the horizon occurs. Thus, from our point of view, $\beta$ acts as a regulator; keeping it small, but finite, is essential for the handling of quantum fluctuations.
   
If we now tune $\Lambda_4$ to non-zero values, we obtain qualitatively different solutions. 
In the simplest case, when there is no chemical potential ($Q=0$), for sufficiently small $\Lambda_4 >0$, and 
$k=1$ (spherical geometry) we recover the de Sitter brane scenario of ref.\ 
\cite{Petkou:2001nk}. As an example, set $M=1$, $\Lambda_4 = 0.5$. For $\beta =0$, we obtain two fixed points
$a = 1.13 $, $2.11$,
outside the outer horizon ($r_+ = 1.03$). As we increase $\beta$ (i.e., $G_5$, or equivalently, decrease $G_4$), 
the larger fixed point increases and the smaller one decreases. After it hits the horizon, the smaller fixed 
point disappears and we only have one fixed point. No fixed points exist inside the horizon. 

\begin{figure}[H]
 \centering

 \subfloat[]{\label{b0l.05-1}
 \includegraphics[width=0.4\textwidth]{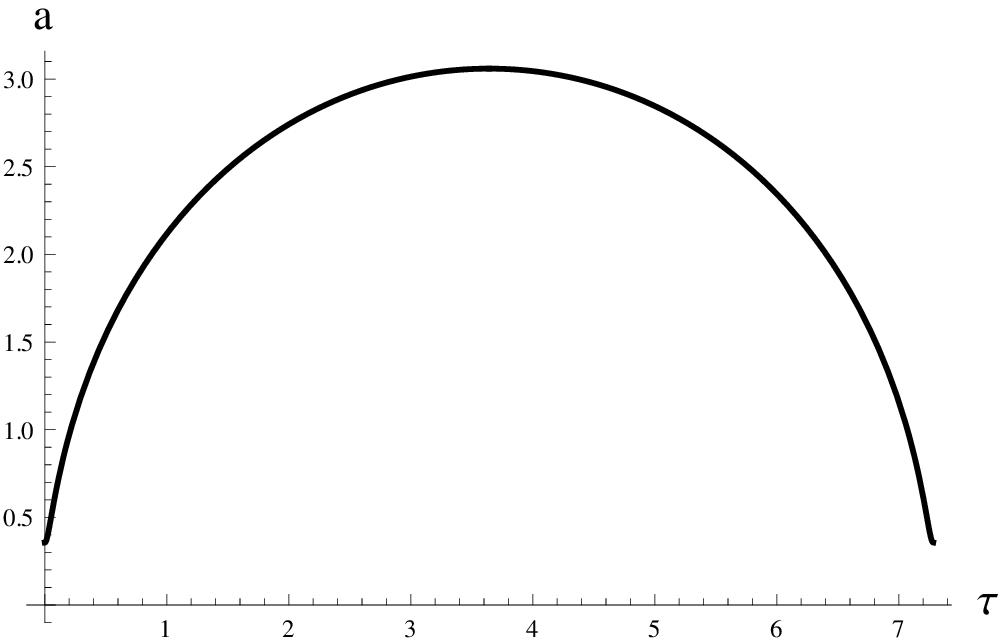}}
 ~~~~ ~~~~ 
 \subfloat[]{\label{b0l.05-2}
 \includegraphics[width=0.4\textwidth]{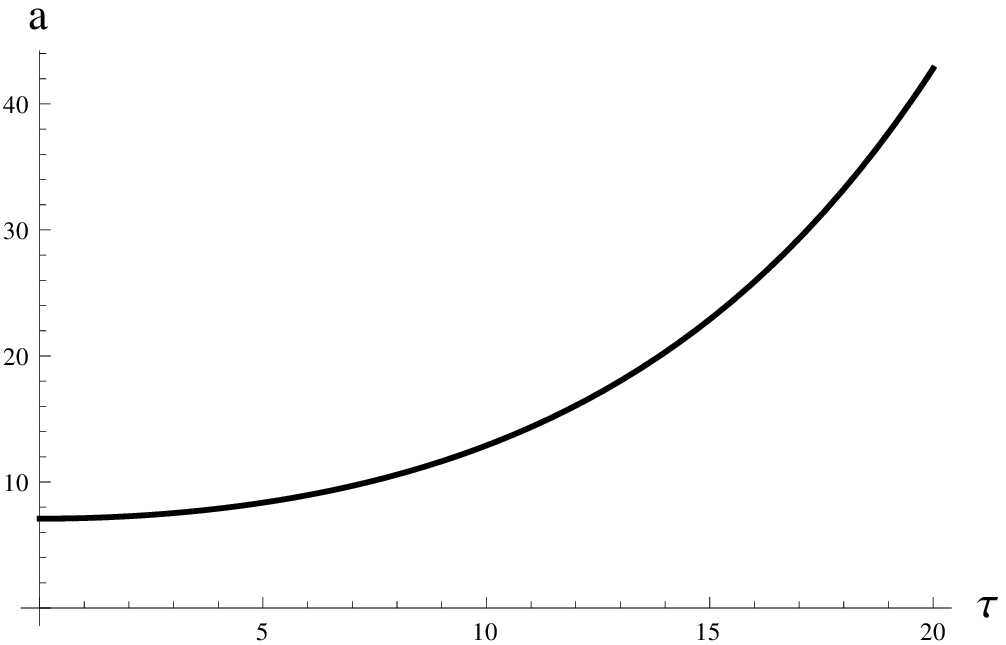}}

\caption{ Plots of $a$ vs $\tau$ for $\beta=0$, $\Lambda_4=0.05$. In (a) we have a bounce. Initial conditions
are chosen as $a(0)= 0.356$. At $\tau=3.642$, $a$ reaches the second fixed point,
$a=3.059$.
In (b) we have an accelerating solution, with initial condition chosen as $a(0)=7.090$.
}
\end{figure}

In the same set up and keeping all other parameters fixed to the afore-mentioned values, if we now turn on
the chemical potential, we obtain one more fixed point away from the outer horizon. For $Q=1$ this is shown in 
figure \ref{fig:1b}. Similarly to the $\Lambda_4 = 0$ case, here we also obtain one bouncing solution with two fixed points, one
inside the inner horizon (figure \ref{bZoom}) and the other outside the outer horizon. This solution for 
$a(\tau)$ is plotted in figure \ref{b0l.05-1}. Additionally, at $a=7.09$ there is another fixed point.
We obtain an accelerating solution from this point (figure \ref{b0l.05-2}). In the region between the first fixed point outside the 
outer horizon ($a=3.06$) and second one at $a=7.09$, the square of the Hubble parameter is negative, hence there is no physical
solution in this region.

\begin{figure}[H]
 \centering
 \includegraphics[width=0.4\textwidth]{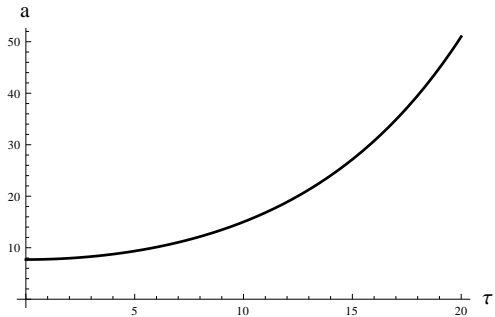}

 \caption{\label{fig:5}Plot of $a$ vs $\tau$ for $\beta=6$, $\Lambda_4=0.05$. 
The initial condition is chosen as $a(0)=7.705$.}

\end{figure}

Comparing the brane world scenario (\ref{V1b}) with the general case, $\beta \ne 0$, we observe that there 
are no qualitative differences in the flat case ($k=0$). In the case of curved horizon (boundary), $k= \pm 1$,
in general one obtains fixed points other than the ones obtained in the brane world scenario. As an example, consider 
the choice of parameters
\begin{equation}\label{eqpara}
 k=+1 \ , \ \ M = 8 \ , \ \ Q = 1 \ , \ \ \Lambda_4 = 0.05 \ , \ \ \beta = 6 \ .
\end{equation}
We have only one fixed point in this case, at $a=7.705$ (figure \ref{fig:1d}). The solution is accelerating as 
shown in figure \ref{fig:5}. There is no bouncing solution for any set of parameters once we go away from 
the special case $\beta=0$.

For $\beta \neq 0$, if we set $\Lambda_4 = 0$, we do not obtain any physical solution. One such situation 
is depicted in figure \ref{fig:1c}. As we see, the square of the Hubble parameter is imaginary for all values of the cosmic scale $a$ in this case. 

\section{Conclusion}\label{sec:6}

In conclusion, we discussed the cosmological evolution derived from a static bulk solution of the field equations with appropriately defined \emph{mixed} boundary conditions using the gravity/gauge theory duality (holography). Such an approach was first discussed in \cite{Apostolopoulos:2008ru}. We extended the results of \cite{Apostolopoulos:2008ru} by considering a boundary hypersurface at arbitrary distance. We calculated the general form of the stress-energy tensor and arrived at a generalized form of the Hubble equation of cosmological evolution. We considered various explicit examples in detail based on an AdS Reissner-Nordstr\"om bulk black hole solution. Interestingly, we obtained the brane-world scenario as a special case, by fine-tuning the parameters of the system, setting $\beta=0$ (eq.\ (\ref{V1ap})). However, keeping $\beta$ small but finite is important in order to avoid scenarios in which the boundary crosses the event horizon from within \cite{Mukherji:2002ft}. Thus, $\beta$ acts as a regulator for such problematic solutions for which quantum fluctuations introduce instabilities \cite{Hovdebo:2003ug}. Moreover, the counterterms one normally introduces to cancel the infinities were shown to have the usual field theoretic interpretation of renormalizing the bare parameters of the system (Newton's constant and the cosmological constant).

It would be interesting to explore the parameter space of the cosmological system further to obtain scenarios of cosmological evolution of interest, such as understanding inflation, and phase transitions in general, in a holographic setting. Various extensions are also possible, such as addition of matter fields on the boundary (without gravity duals). Also, anisotropic cosmologies are possible from a static bulk background, if the boundary hypersurface is chosen with a different geometry than the horizon (e.g., flat boundary ($k=0$) in a bulk black hole background of spherical horizon ($k=+1$)). Work in this direction is in progress \cite{bibfuture}.

\section*{Acknowledgments}

We are grateful to Sudipta Mukherji for discussions. G.~S.~was
supported in part by the US Department of Energy under Grant No.\
DE-FG05-91ER40627.


\end{document}